\documentstyle[11pt]{article}

\def\kon#1#2{\vbox{\halign{##&&##\cr\lower4pt
\hbox{$\scriptscriptstyle\vert$}\hrulefill &\hrulefill\lower4pt
\hbox{$\scriptscriptstyle\vert$}\cr $#1$&$#2$\cr}}}

\def\fii{\varphi}
\def\al{\alpha}
\def\be{\beta}

\def\ro{\rho}

\def\d{\partial}
\def\=d{\,{\buildrel\rm def\over =}\,}

\def\sqr#1#2{{\vcenter{\vbox{\hrule height.#2pt\hbox{\vrule width.
#2pt height#1pt \kern#1pt \vrule width.#2pt}\hrule height.#2pt}}}}

\def\la{\lambda}

\def\te{\vartheta}

\begin{document}

\title{FROM MASSIVE GRAVITY
TO DARK MATTER DENSITY II}
\author{\\G. Scharf
\footnote{e-mail: scharf@physik.unizh.ch}
\\ Institut f\"ur Theoretische Physik, 
\\ Universit\"at Z\"urich, 
\\ Winterthurerstr. 190 , CH-8057 Z\"urich, Switzerland}

\date{}

\maketitle\vskip 3cm

\begin{abstract} As previously observed the massless limit 
of massive gravity leads to a modification of general relativity. 
Here we study spherically symmetric solutions of the modified 
field equations which contain normal matter together with a
dark energy density. If the dark density profile is
assumed to be known, the whole problem is reduced to a linear
first order differential equation which can be solved by
quadratures.

\end{abstract}

\newpage

\section{Introduction}

In a previous paper with the same title [1] we have made the
thousand and first proposal to explain dark matter density.
In contrast to the other thousand proposals we came to this subject
by accident, that means not by looking for an explanation of dark
matter. As the title indicates we have studied massive quantum gravity
which is gravity with a massive graviton. To have a massive spin-2
gauge theory the so-called vector graviton field $v_\la$ is indispensable.
The crucial observation was that in the limit $m\to 0$ of vanishing
graviton mass this field $v_\la$ does not decouple from the symmetric
tensor field $h^{\mu\nu}$ which is equivalent to Einstein's $g^{\mu\nu}$.
In the classical limit $v_\la$ acts as four scalar fields $v_n$. As a
consequence the massless limit of massive gravity is different from general
relativity. There remains the additional coupling to the four (now
massless) scalar fields $v_n$. In the resulting modified Einstein's
equations this gives additional terms in the energy - momentum
tensor. In the 00-component the new contribution looks as if it comes
from a dark matter density, but there are also peculiar modifications
of the pressure components in the $jj$-equations.

The paper is organized as follows. In the next section we review the
modified general relativistic equations, for their derivation we refer to [1].
We then consider static spherically symmetric solutions including normal
matter which is described by a mass density $q(r)$ and an isotropic pressure $p(r)$.
By expanding the solution for large distance $r$ we get already an important
result: the dark density profile must decrease as $1/r^4$ as already found 
in [1] for the solution without matter. This tail contradicts the widely
discussed Navarro - Frenk - White (NFW) profile [2] [3]
$$\ro (r)={\ro_s\over (r/r_s)(1+(r/r_s))^2}\eqno(1.1)$$
which has a $1/r^3$ tail. However, this profile must be modified for large
$r$ anyway to get a finite total mass for the dark halo. A second conclusion
can be drawn from the expansion around $r=0$. In general relativity there
exists the inner Schwarzschild solution which is finite in all quantities
including mass density and pressure. This is of course also a solution of
our modified theory with vanishing dark density. We can now test whether there
is a corresponding {\it finite} solution with a dark density different from zero.
The answer is no. This may explain the fact that a single star like the sun
which is described by this finite solution does not have a dark halo. In order to
describe dark halos we have to study solutions with some singularity for small $r$.
In [1] we have investigated the vaccum solutions, here we start the analysis of
solutions with normal matter.

The modified general relativity has the following nice property. To specify a solution
one quantity must be given which usually is the mass density $q(r)$ of normal matter.
In this case one has to solve non-linear differential equations. Alternatively, if
the dark profile $\ro(r)$ is considered to be known, the remaining differential
equation is linear and first order. It can be solved by quadratures. This is
a machine which produces solutions at low cost. In this way
we construct a simple singular solution in section 3. This solution with a dark
halo has a singularity at some finite radius $r_s$. There is no horizon at $r>r_s$
so that the singularity is a naked one. 

\section{Massless limit of massive gravity}

The basic classical field equations which follow from massive gravity in the limit
of vanishing graviton mass [1] are the modified Einstein equations
$$R_{\mu\nu}-{1\over 2}g_{\mu\nu}R-{8\pi G\over c^4}t_{\mu\nu}=
{16\pi G\over c^3}\Bigl(-\d_\mu v_n\d_\nu v^n+
+{1\over 2}g_{\mu\nu}g^{\al\be}\d_\al v_n\d_\be v^n\Bigl).\eqno(2.1)$$
together with the Laplace-Beltrami wave equation for the four scalar fields
$v^n, n=0,1,2,3$
$$\d_\al(\sqrt{-g}g^{\al\be}\d_\be v^n)=0.\eqno(2.2)$$
As discussed in [1] the Latin index $n$ of $v^n$ is raised and lowered
with the Minkowski tensor ${\rm diag} (1,-1,-1,-1)$, in contrast to the
Greek indices which are changed with the metric tensor $g^{\mu\nu}$.
Since upon $v^n$ we act by partial derivatives not by covariant ones,
this means that the $v_n$ are four scalar fields, indeed. The new contribution
on the right-hand side of (2.1) is the possible origin of the dark density.

We want to study static spherically symmetric solutions of these
field equations. As in [1] we write the metric as
$$ds^2=e^\nu c^2dt^2-e^\la dr^2-r^2(d\te^2+\sin^2\te\d\fii^2)\eqno(2.3)$$
where $\nu$ and $\la$ are functions of $r$ only. We take the coordinates
$x^0=ct$, $x^1=r$, $x^2=\te$, $x^3=\fii$ such that
$$g_{00}=e^\nu,\quad g_{11}=-e^\la$$
$$g_{22}=-r^2,\quad g_{33}=-r^2\sin^2\te\eqno(2.4)$$
and zero otherwise. The components with upper indices are the inverse of
this. The determinant comes out to be
$$g={\rm det}g_{\mu\nu}=-e^{\nu+\la}r^4\sin^2\te.\eqno(2.5)$$

The energy momentum tensor in (2.1) is assumed in the simple
form
$$t_\al^{\>\be}={\rm diag}(qc^2,-p,-p,-p)\eqno(2.6)$$
where $q(r)$ is the ordinary mass density and $p(r)$ is 
an isotropic pressure.
Now the following modified radial Einstein equations must be solved
$$e^{-\la}\Bigl({\la'\over r}-{1\over r^2}\Bigl)+{1\over r^2}
={8\pi G\over c^3}(q(r)c+w_0(r))\eqno(2.7)$$
$$e^{-\la}\Bigl({\nu'\over r}+{1\over r^2}\Bigl)-{1\over r^2}
={8\pi G\over c^3}({p(r)\over c}-w_0(r))\eqno(2.8)$$
$$e^{-\la}\Bigl({\nu''\over 2}+{\nu^{\prime 2}\over 4}- 
{\nu'\la'\over 4}-{\la'\over 2r}+{\nu'\over 2r}\Bigl)
={8\pi G\over c^3}({p(r)\over c}+w_0(r)),\eqno(2.9)$$
where $w_0$ is the contribution of the four scalar fields $v^n$.
As discussed in [1] in the spherically symmetric case this is of the form 
$$w_0={\ro_0\over r^4}e^{-\nu(r)}.\eqno(2.10)$$

First by suitable combination we simplify the equations. Adding (2.7) to
(2.8) we get
$${e^{-\la}\over r}(\la'+\nu')=g(qc+{p\over c})\eqno(2.11)$$
where
$$g={8\pi G\over c^3}.\eqno(2.12)$$
Eliminating $p$ from (2.8) and (2.9) we obtain
$$\nu''=2e^\la\Bigl(-{1\over r^2}+2g{w_0\over c}\Bigl)+{\nu'\la'\over 2}
-{\nu^{\prime 2}\over 2}+{\la'+\nu'\over r}+{2\over r^2}.
\eqno(2.13)$$
Next we differentiate (2.8) with respect to $r$ and use (2.13):
$$g\Bigl({p'\over c^2}-{w'_0\over c}\Bigl)=-e^{-\la}{\nu'\over 2r}
(\la'+\nu')+4{g\over cr}w_0.\eqno(2.14)$$
Substituting (2.11) inhere we finally arrive at
$${p'\over c^2}=-{\nu'\over 2}\Bigl(q+{p\over c^2}+{2\over c}w_0
\Bigl)\eqno(2.15)$$
where (2.10) has been used. This differential equation for the pressure
will be used instead of the second order equation (2.9) in the
following.

To study the solution for large $r$ where it should approach flat space,
we set up an expansion in powers of $1/r$:
$$e^{-\la}=1-{a_1\over r}+{a_2\over r^2}+\ldots\eqno(2.16)$$
$$\nu={b_1\over r}+{b_2\over r^2}+\ldots\eqno(2.17)$$
$$q={q_4\over r^4}+{q_5\over r^5}+\ldots\eqno(2.18)$$ 
$$p={p_5\over r^5}+{p_6\over r^6}+\ldots\eqno(2.19)$$
We find 
$$a_1=-b_1\eqno(2.20)$$
$$a_2=g(q_4+\ro_1),\quad \ro_1={\ro_0\over c}\eqno(2.21)$$
$$q_4={1\over g}(4b_2+b_1^2)-2\ro_1\eqno(2.22)$$
$${p_5\over c^2}=-{b_1\over 10}(q_4+2\ro_1)=-{b_1\over 10g}(4b_2+b_1^2).\eqno(2.23)$$
If the dark density $w_0(r)$ is given (2.10), i.e. $\ro_0$ and $\nu(r)$ are known,
we have a unique solution. The equation of state of the normal matter is then
determined.

In ordinary general relativity there exists a finite inner solution which
can be expanded around $r=0$ in the form
$$e^{-\la}=1+a_1r+a_2r^2+a_3r^3+\ldots$$
$$\nu =b_0+b_1+b_2r^2+b_3r^3+\ldots\eqno(2.24)$$
$$qc=q_0+q_1r+q_2r^2+\ldots$$
$${p\over c}=p_0+p_1r+p_2r^2+\ldots.$$
Substituting this into (2.7-9) with $w_0=0$ we find
$$a_1=0,\quad a_2=-{g\over 3}q_0,\quad a_3=-{g\over 4}q_1\eqno(2.25)$$
$$b_1=0,\quad b_2={g\over 2}(p_0+{q_0\over 3}),\quad b_3={g\over 12}q_1\eqno(2.26)$$
$$p_1=0,\quad p_2=-{g\over 4}(p_0+{q_0\over 3})(p_0+q_0)\eqno(2.27)$$
$$p_3=-{g\over 6}¢p_0+{7\over 12}q_0q_1).$$
Now we can check whether this finite solution has a counterpart with non-vanishing
$w_0$. Expanding (2.10)
$$w_0(r)={\ro_0\over r^4}e^{-\nu}=\ro_0e^{-b_0}\Bigl({1\over r^4}-{b_2\over 
r^2}+\ldots\Bigl),$$
we easily see from (2.7) that $\ro_0$ must vanish. A solution finite at $r=0$
cannot have a dark halo. Therefore we have to look for singular solutions.

\section{A singular solution}

To construct a singular solution with non-vanishing dark density $w_0(r)$
we proceed as follows. We eliminate $p(r)$ in the two equations (2.8) (2.9):
$$-\la'e^{-\la}\Bigl({\nu'\over 4}+{1\over 2r}\Bigl)=e^{-\la}\Bigl(-{\nu'' 
\over 2}-{\nu^{\prime 2}\over 4}+{\nu'\over 2r}+{1\over r^2}\Bigl)-$$
$$-{1\over r^2}+2gw_0.\eqno(3.1)$$
Introducing
$$y=e^{-\la}\eqno(3.2)$$
we have obtained a linear first order differential equation for $y(r)$:
$$f_1(r)y'(r)=f_2(r)y+f_3(r)\eqno(3.3)$$
where
$$f_1={\nu'\over 4}+{1\over 2r}\eqno(3.4)$$
$$f_2=-{\nu''\over 2}-{\nu^{\prime 2}\over 4}+{\nu'\over 2r}+{1\over r^2}
\eqno(3.5)$$
$$f_3=-{1\over r^2}+2gw_0.\eqno(3.6)$$
Considering the dark density $w_0(r)$ (2.10) as given we can compute
$$\nu(r)=\log\ro_0-\log w_0-4\log r.\eqno(3.7)$$
This enables us to express the coefficients $f_1, f_2, f_3$ in terms of $w_0$:
$$f_1=-{w'_0\over 4w_0}-{1\over 2r}\eqno(3.8)$$
$$f_2={w''_0\over 2w_0}-{3\over 4}{w_0^{\prime 2}\over w_0^2}-{5\over 2}
{w'_0\over rw_0}-{7\over r^2}\eqno(3.9)$$
$$f_3=-{1\over r^2}+2gw_0.\eqno(3.10)$$
The remaining equation (2.7) then determines the mass density $q(r)$ of
the normal matter.

The linear equation (3.3) can be simply solved by quadratures, so we have
full control of the solution. The general solution $y(r)$ is the sum of a
particular solution $y_1(r)$ of the inhomogeneous equation (3.3) plus a
solution $y_0(r)$ of the homogeneous equation. We will soon realize that
the solution is uniquely fixed by requiring that it approaches flat space
for $r\to\infty$. To see this we set up an expansion of the form
$$y=1-{a_1\over r}+{a_2\over r^2}+\ldots\eqno(3.11)$$
In agreement with the expansion (2.16-19) above the dark density (2.10) must 
start as follows
$$w_0={\ro_0\over r^4}+{d_1\over r^5}+{d_2\over r^6}+\ldots\eqno(3.12)$$
A $1/r^3$ tail is in contradiction with the flat space asymptotic
$y\to 1$ for $r\to\infty$. From (3.12) and (3.8) we find
$$f_1={1\over 2r}+{d_1\over 4\ro_0}{1\over r^2}+\Bigl({d_2\over 2\ro_0}
-{d_1^2\over 4\ro_0^2}\Bigl){1\over r^3}+\ldots\eqno(3.13)$$
and similarly
$$f_2={1\over r^2}+{3d_1\over 2\ro_0}{1\over r^3}+\Bigl({4d_2\over\ro_0}
-{9d_1^2\over 4\ro_0^2}\Bigl){1\over r^4}+\ldots\eqno(3.14)$$
$$f_3=-{1\over r^2}+2g\Bigl({\ro_0\over r^4}+{d_1\over r^5}+{d_2\over 
r^6}+\ldots\Bigl).\eqno(3.15)$$
Substituting all this into (3.3) we can determine the coefficients in (3.11).
The result is
$$y=1-{d_1\over\ro_0r}+\Bigl({2d_1^2\over\ro_0^2}-{2d_2\over\ro_0}-g\ro_0\Bigl)
{1\over r^2}+\ldots\eqno(3.16)$$

To carry out all integrations in terms of elementary functions we choose the
following simple dark density profile
$$w_0={w_1\over 1+(r/r_s)^4}.\eqno(3.17)$$
It contains two parameters $w_1$ and $r_s$ as most phenomenological dark matter
profiles in the literature and has the correct asymptotic behavior. First we
want to calculate the solution $y_0(r)$ of the homogeneous equation
$$f_1y'_0=f_2y_0,\eqno(3.18)$$
which is given by
$$y_0(r)=\exp\int\limits^r{f_2\over f_1}dx.\eqno(3.19)$$
Now from (3.8) and (3.17) we get
$$f_1={r^4-r_s^4\over 2r(r^4+r_s^4)}.\eqno(3.20)$$
In view of the integral in (3.19) it is convenient to use the original
definitions (3.4) (3.5) which yield
$${f_2\over f_1}=-2{f'_1\over f_1}-\nu'+{4\over r}-{2\over r^2f_1}.\eqno(3.21)$$
In our special case (3.17) this is equal to
$${f_2\over f_1}=-2{f'_1\over f_1}-\nu'-{8r_s^4\over r(r^4-r_s^4)}.\eqno(3.22)$$
This can be easily integrated and exponentiated using (2.10)
$$e^{-\nu}=r^4{w_0\over\ro_0}.$$
In this way we find the following solution (3.19)
$$y_0(r)={\rm const.}{r^{14}(r^4+r_s^4)\over (r^4-r_s^4)^4}.\eqno(3.23)$$
The solution of the linear homogeneous equation contains a free prefactor,
of course. For $r\to\infty$ $y_0(r)$ increases as $r^2$ which is in conflict
with the asymptotic flatness $y\sim$ const. Consequently, the prefactor of $y_0$
has to be zero so that we get a unique solution of our problem as in the expansion
(2.16) in the last section.

From $y_0(r)$ on finds the solution $y_1(r)$ of the inhomogeneous equation by the 
well-known method of variation of the constants. One makes the ansatz
$$y_1(r)=a(r)y_0(r),\eqno(3.24)$$
where $y_0$ is given by (3.23) with prefactor 1. Substituting this into (3.3) one
obtains $a(r)$ by another integration:
$$a(r)=\int\limits^r {f_3(x)\over f_1(x)y_0(x)}dx,\eqno(3.25)$$
where a constant of integration, i.e. the lower limit of the integral is free.
The integration of the rational function (3.25) is elementary. The final result is
$$a(r)={1\over r^2}-{r_s^4\over r^6}+{3\over 5}{r_s^8\over r^{10}}-
{r_s^{12}\over 14r^{14}}+$$
$$+4gw_1\Bigl[-2\log\Bigl(1+{r_s^4\over r^4}\Bigl)+{7\over 4}{r_s^4\over r^4}
-{1\over 2}{r_s^8\over r^8}+{1\over 12}{r_s^{12}\over r^{12}}\Bigl].\eqno(3.26)$$
Substituting this back into (3.24) we get the desired unique solution which we
denote by $y(r)$ again:
$$y(r)={r^4+r_s^4\over (r^4-r_s^4)^4}\Bigl\{r^{12}-r_s^4r^8+{3\over 5}r_s^8r^4
-{1\over 7}r_s^{12}+$$
$$+gw_1\Bigl[-8r^{14}\log(1+{r_s^4\over r^4})+7r_s^4r^{10}-2r_s^8r^6+{1\over 3}
r_s^{12}r^2\Bigl]\Bigl\}.\eqno(3.27)$$

The solution $y(r)$ has the correct asymptotic behavior $y\to 1$ for
$r\to\infty$. However, there is a singularity at $r=r_s$. We ask whether this
singularity is hidden by a horizon. A horizon corresponds to $y=\exp (-\la)=0$.
So we look for a zero of the curly bracket in (3.27). If we insert realistic
numbers for the dark matter density $w_1$ as discussed in the next section,
we find that the second line in (3.27) is completely neglegible compared to 
the first one due to the smallness of $g$. The polynomial
$$r^{12}-r_s^4r^8+{3\over 5}r_s^8r^4-{1\over 7}r_s^{12}$$
has only one real zero $r_0<r_s$. Consequently the singularity can be seen
from the outside, that means it is naked.

Finally the mass density $q(r)$ of normal matter is obtained from (2.7)
according to
$$cq(r)={1\over g}\Bigl(-{y'\over r}-{y\over r^2}+{1\over r^2}\Bigl)
-w_0.\eqno(3.28)$$
This also becomes singular at $r=r_s$.

\section{Comparison with observations}

At first side it is questionable whether our solution with a spherically symmetric
density distribution $q(r)$ of the normal matter can be compared to observations
of spiral galaxies. However, since the dark halo is approximately spherically
symmetric, at least the gas component in the outer part of the halo can be assumed
to be also spherically symmetric [4]. If we chose a galaxy which is dark matter
dominated where the normal matter is only 10 percent of the total mass, say, then
a spherically symmetric description is not a bad approximation. The spiral galaxy
M33 is a good example for our purpose. Edvige Corbelli [5] has obtained good fits
to the data which give (via rotation curves) the density profiles of dark and
visible matter. The data are compatible with very different forms of the dark
density profile. Beside the NFW profile (1.1) Corbelli considers the isothermal
profile
$$\ro(r)={\ro_i\over (1+(r/r_i))^2}\eqno(4.1)$$
and the Burkert profile [6]
$$\ro (r)={\ro_B\over (1+r/r_B)(1+(r/r_B)^2)},\eqno(4.2)$$
which both have a central constant - density core like our profile (3.17).
The data are not much constraining the profile in the central region, and in
the tail region $r > 17$ kpc there are no data. Consequently it is well
possible to fit our profile (3.17) to the best fit of Corbelli between, say,
4 and 14 kpc. However, the resulting singular radius $r_s$ comes out to be
around 9 kpc. This is too big. We could except a value $r_s < 0.5$ kpc
which  is the radius of the bright ``nucleus''[5], where something unknown
is going on.

A similar conclusion can be drawn from the analysis of the following 3-parameter
profile
$$w_0={w_1\over r^2(1-a_1r+a_2r^2)}.\eqno(4.3)$$
Then the solution of the homogeneous equation (3.18) is given by
$$y_0(r)={r^{-2-16a_2/a_1^2}(1-a_1r+a_2r^2)\over \vert 2a_2r-a_1\vert^{
16a_2/a_1^2}}\exp\Bigl(-{8\over a_1r}\Bigl).\eqno(4.4)$$
This produces a singularity at $r_s=a_1/2a_2$. Again fitting (4.3) to
Corbelli's result for the dark matter profile we get $r_s$ around 10 kpc
which is unacceptable.

The occurrence of a singularity seems to be a general feature of the simple
equation (3.3). It is a consequence of the integrability of the problem which
is due to the assumption of an isotropic pressure $p(r)$ in (2.6). On the other
hand the dark contributions $\mp w_0(r)$ in (2.8) and (2.9) appear with different
signs. Since the normal matter is probably following the dark in a certain way,
the pressure components $p_j(r)$ in (2.8) and (2.9) should not be the same.
We shall investigate this more general situation in the next part of this series.

\end{document}